\newcommand{\be}{\begin{equation}}\newcommand{\ee}{\end{equation}}
\newcommand{\bea}{\begin{eqnarray}}\newcommand{\eea}{\end{eqnarray}}
\newcommand{\nn}{\nonumber}\newcommand{\p}[1]{(\ref{#1})}
 \newcommand{\lb}[1]{\label{#1}}
 \newcommand\q{\quad}
\newcommand\qq{\quad\quad}
\newcommand\s{\scriptscriptstyle}

\def\a{\alpha}

\def\b{\beta}

\def\g{\gamma}

\def\d{\delta}

\def\eps{\epsilon}

\def\ve{\varepsilon}

\def\j{\psi} 
  
\def\l{\lambda}
 \def\th{\theta}  \def\bt{\bar\theta}

\def\r{\rho}

\def\z{\zeta}

\def\J{\Psi}

\def\L{\Lambda}

\def\pa{\partial}

\newcommand\Tr{\mbox{Tr}\,}

\newcommand\tooS{(\theta^{(0,0)})^2}

\newcommand\ab{{\alpha\beta}}

\newcommand\A{{\s A}}

\newcommand{\Dpp}{\cD^{(1,1)}}
\newcommand{\Dpm}{\cD^{(1,-1)}}

\newcommand{\DPo}{\cD^{(2,0)}}

\newcommand{\DoP}{\cD^{(0,2)}}
\newcommand{\DoM}{\cD^{(0,-2)}}
\newcommand{\lpp}{\l^{(1,1)}}
\newcommand{\lpm}{\l^{(1,-1)}}
\newcommand{\lpo}{\l^{(2,0)}}

\newcommand{\Vpp}{V^{(1,1)}}
\newcommand{\Vpm}{V^{(1,-1)}}

\newcommand\cD{{\cal D}}

\def\sfrac#1#2{{\textstyle\frac{#1}{#2}}}

\documentclass[12pt]{article}
\usepackage{amsfonts}
\begin{document}

\begin{center}
{\bf     CHERN-SIMONS $D=3, N=6$ SUPERFIELD THEORY}\\
\vspace{0.5cm} {\it
  B.M. Zupnik\\
 Bogoliubov Laboratory of Theoretical Physics, JINR, Dubna,
  Moscow Region, 141980, Russia; E-mail: zupnik@theor.jinr.ru}
\vspace{1cm}
\end{center}
\begin{abstract}
We construct  the $D{=}3, N{=}5$ harmonic superspace using  the 
SO(5)/U(1)$\times$U(1) harmonics. Three gauge harmonic superfields 
satisfy the off-shell constraints of the Grassmann and harmonic 
analyticities. The corresponding component supermultiplet contains 
the gauge field $A_m$ and an infinite number of bosonic and fermionic 
fields with the SO(5) vector indices arising from decompositions of 
gauge superfields in harmonics and Grassmann coordinates. The 
nonabelian superfield Chern-Simons action is invariant with respect 
to the $N{=}6$ superconformal supersymmetry realized on the $N{=}5$ 
superfields. The component Lagrangian contains the Chern-Simons 
interaction of $A_m$ and an infinite number of bilinear and trilinear 
interactions of auxiliary fields. The fermionic and bosonic auxiliary 
fields from the infinite $N{=}5$ multiplet vanish on-shell.
\end{abstract}

\setcounter{equation}0
\section{Introduction}

Supersymmetric extensions of the three-dimensional Chern-Simons 
(CS) theory were discussed in refs. \cite{Si}-\cite{Schw}. The 
$N{=}1$ CS theory of the spinor gauge superfield  \cite{Si,Scho} 
was constructed in the $D{=}3, N{=}1$ superspace with real 
coordinates $x^m, \th^\a$, where $m=0, 1, 2$ is the 3D vector 
index and $\a=1, 2$ is the SL(2,R) spinor index. The $N{=}1$ CS 
action can be interpreted as the superspace integral of the 
Chern-Simons superform $dA+\sfrac23A^3$ in the framework of our 
theory of superfield integral forms \cite{ZP1}-\cite{Z6}.

The abelian $N{=}2$ CS action was first constructed in the $D{=}3, 
N{=}1$ superspace \cite{Si}. The nonabelian $N{=}2$ CS action was 
considered in the $D{=}3, N{=}2$ superspace in terms of the 
Hermitian superfield $V(x^m,\th^\a,\bt^\a)$ (prepotential) 
\cite{ZP1,Iv,NG}, where $\th^\a$ and $\bt^\a$ are the complex 
conjugated $N{=}2$ spinor coordinates. The corresponding 
component-field Lagrangian includes the bosonic CS term and the 
bilinear terms with fermionic and scalar fields without 
derivatives. The unusual dualized form of the $N{=}2$ CS 
Lagrangian contains the second vector field instead of the 
scalar field\cite{NG}.

The $D{=}3, N{=}3$ CS theory was first analyzed by the 
harmonic-superspace method \cite{ZK,Z3}. Note that the 
off-shell $N{=}3$ and $N{=}4$ vector supermultiplets are identical 
\cite{Z5}; however, the superfield CS action is invariant with 
respect to the $N=3$ supersymmetry only. Nevetheless, the $N{=}3$ 
CS equations of motion are covariant under the 4th 
supersymmetry. The field-component form of the $N{=}3$ CS 
Lagrangian was studied in \cite{KL}.

The off-shell $D{=}3, N{=}6$ SYM theory arises by a dimensional 
reduction of the $D{=}4, N{=}3$ SYM theory in the 
SU(3)/U(1)$\times$U(1) harmonic superspace \cite{GIOS}. Three basic 
prepotentials of the $D{=}3, N{=}6$ gauge theory contain an infinite 
number of  auxiliary fields with the SU(3) indices, and a coupling 
constant of this model has a dimension 1/2. We do not know how to 
construct the $D{=}3, N{=}6$ CS theory from these gauge harmonic 
superfields. Note that  the SU(3)/U(1)$\times$U(1) analytic harmonic 
superspace  has the integration measure of  dimension 1 in the 
case $D{=}3$.

In this paper, we consider the  simple  $D{=}3, N{=}5$ superspace 
which cannot be obtained by a dimensional reduction of the even 
coordinate from any 4D superspace. The corresponding harmonic 
superspace  using the SO(5)/U(1)$\times$U(1) harmonics is 
discussed in Section 2. The Grassmann-analytic $D{=}3, N{=}5$ 
superfields depend on 6 spinor coordinates, so the 
analytic-superspace integral measure has a zero dimension. It 
is shown that five harmonic derivatives preserve the Grassmann 
analyticity.

In Section 3, we consider five basic  gauge superfields in the 
$D{=}3, N{=}5$ analytic superspace and the corresponding  gauge 
group with analytic superfield parameters. To simplify  
the superfield formalism of the theory, one can introduce  
additional off-shell harmonic-analyticity constraints for the 
gauge-group parameters and gauge superfields. These harmonic 
constraints yield additional reality conditions for components 
of superfields. In this convenient representation, two gauge 
superfields vanish, and we use only three basic gauge 
superfields (prepotentials). The Chern-Simons superfield action 
can be constructed from these $D{=}3, N{=}5$ gauge superfields. We 
show that this CS superfield action is invariant with respect to the 
$D{=}3, N{=}6$ superconformal supersymmetry transformations realized 
on the $N{=}5$ superfields. The superfield gauge equations of motion 
have only pure gauge solutions, by analogy with the $N{=}1, 2, 3$ 
superfield CS theories.

The field-component structure of our $D{=}3, N{=}6$ Chern-Simons 
model is analyzed in Section 4. In the abelian case, the basic  
gauge superfield includes the gauge field $A_m$ and the fermion 
field  $\psi_\a$ in the SO(5) invariant sector, and an infinite 
number of fermionic and bosonic fields with the SO(5) vector or 
tensor indices. The component  Lagrangian contains the Chern-Simons 
term for $A_m$ and  the simple bilinear and trilinear interactions 
of other fermionic and bosonic fields. The field strength of the gauge 
field and all other fields vanish on-shell.

The preliminary version of the $D{=}3, N{=}5$ harmonic-superspace 
gauge theory without the harmonic-analyticity conditions was 
presented in our talk \cite{Z7}. This model describes the 
interaction of the $N{=}5$ Chern-Simons multiplet with some 
unusual matter fields.

\setcounter{equation}0
\section{$D=3, N=5$ harmonic superspace}

The CB-representation of the $D{=}3, N{=}5$ superspace uses three 
real even coordinates $x^m\q (m=0, 1, 2)$ and five 
two-component odd  coordinates $\th^\a_a$, where $\a=1, 2$ is 
the spinor index of the group SL(2,R) and $a=1, 2,\ldots 5$ is 
the vector index of the automorphism group SO(5). We use the 
real traceless or symmetric representations of the 3D $g_m$ 
matrices 
\bea 
&&(\g_m)^\ab=\ve^{\a\r}(\g_m)^\b_\r=(\g_m)^{\b\a},\q
(\g^m)_\ab(\g_m)^{\r\g}=\d_\a^\r\d_\b^\g+\d_\b^\r\d_\a^\g\nn\\
&&(\g_m\g_n)^\b_\a=(\g_m)_\a^\r(\g_n)_\r^\b=-(\g_m)_{\a\r}(\g_n)^{\r\b} 
=-\eta_{mn}\d^\b_\a+\ve_{mnp}(\g^p)^\b_\a. 
\eea 
where $\eta_{mn}=\mbox{diag}(1,-1,-1)$ is the 3D Minkowski metric and 
$\ve_{mnp}$ is the antisymmetric symbol.

One can consider the bispinor representation of the 3D 
coordinates and derivatives 
\bea 
&&x^\ab=(\g_m)^\ab x^m,\q \pa_\ab=(\g^m)_\ab \pa_m. 
\eea

The $N{=}5$ CB spinor derivatives are 
\bea 
&&D_{a\a}=\pa_{a\a}+i\th^\b_a\pa_\ab,\q\pa_{a\a}\th^\b_b=\d_{ab}\d^\b_\a. 
\eea

The $N{=}5$   supersymmetry transformations are 
\be 
\d_\eps x^m=-i\eps_a^\a(\g^m)_\ab\th^\b_a,\q \d_\eps \th^\a_a=\eps^\a_a. 
\ee

We shall use the SO(5)/U(1)$\times$U(1) vector harmonics  
defined via the components of the real orthogonal 5$\times$5 
matrix 
\be 
U^K_a=\left(U^{(1,1)}_a, U^{(1,-1)}_a, U^{(0,0)}_a, 
U^{(-1,1)}_a, U^{(-1,-1)}_a\right) 
\ee 
where $a$ is the SO(5) vector index and the index $K=1, 2,\ldots 5$ 
corresponds to given combinations of the U(1)$\times$U(1) charges. 
These harmonics satisfy the following conditions: 
\bea
&&U^K_aU^L_a=g^{KL}=g^{LK},\q g^{KL}U^K_aU^L_b=\d_{ab},\lb{Ubasic}\\
&& g^{15}=g^{24}=g^{33}=1,\q 
g^{11}=g^{12}=\cdots=g^{45}=g^{55}=0,\nn 
\eea 
where $g^{LK}$ is the antidiagonal symmetric constant metric in the  
space of charged indices.

Let us introduce the following harmonic derivatives: 
\bea
&&\pa^{KL}=U^K_bg^{LM}\frac{\pa}{\pa U^M_b}-U^L_bg^{KM}\frac{\pa}{\pa U^M_b}
=-\pa^{LK},\\
&&[\pa^{IJ},\pa^{KL}]=g^{JK}\pa^{IL}+g^{IL}\pa^{JK}-g^{IK}\pa^{JL}-g^{JL}
\pa^{IK}, 
\eea 
which satisfy the commutation relations of the Lie algebra SO(5). We will 
mainly use the five harmonic derivatives and the corresponding U(1)$\times$U(1)
notation
\bea 
&&\pa^{12}=\pa^{(2,0)}=U^{(1,1)}_b\pa/\pa 
U^{(-1,1)}_b-U^{(1,-1)}_b\pa/\pa U^{(-1,-1)}_b,\nn\\
&&\pa^{13}=\pa^{(1,1)}=U^{(1,1)}_b\pa/\pa 
U^{(0,0)}_b-U^{(0,0)}_b\pa/\pa U^{(-1,-1)}_b,\nn\\
&&\pa^{23}=\pa^{(1,-1)}=U^{(1,-1)}_b\pa/\pa 
U^{(0,0)}_b-U^{(0,0)}_b\pa/\pa U^{(-1,1)}_b,\nn\\
&&\pa^{14}=\pa^{(0,2)}=U^{(1,1)}_b\pa/\pa
U^{(1,-1)}_b-U^{(-1,1)}_b\pa/\pa U^{(-1,-1)}_b,\nn\\
&&\pa^{25}=\pa^{(0,-2)}=U^{(1,-1)}_b\pa/\pa 
U^{(1,1)}_b-U^{(-1,-1)}_b\pa/\pa U^{(-1,1)}_b. \nn 
\eea

The Cartan  charges of two U(1) groups are described by the 
neutral harmonic derivatives 
\bea 
\pa^0_1=\pa^{15}+\pa^{24},\q \pa^0_1U^{(p,q)}_a=p\,U^{(p,q)}_a,
\q\pa^0_2=\pa^{15}-\pa^{24},\q \pa^0_2U^{(p,q)}_a=q\,U^{(p,q)}_a. 
\eea

The harmonic integral has the following simple properties: 
\be 
\int dU=1,\q \int dU\,U^{(p,q)}_aU^{(-r,-s)}_b=\sfrac15\d_{ab}\d_{pr}\d_{qs}.
\lb{Uint} 
\ee

Let us define the harmonic projections of the $N{=}5$ Grassmann 
coordinates 
\bea 
&&\th^K_\a=\th_{a\a}U^K_a=(\th^{(1,1)}_\a, \th^{(1,-1)}_\a, \th^{(0,0)}_\a, 
\th^{(-1,1)}_\a, \th^{(-1,-1)}_\a). 
\eea

Using the  harmonic-superspace  method  one can  define the coordinates 
of the $N{=}5$ analytic superspace with only three spinor coordinates 
\bea
&&\zeta=( x^m_\A, \th^{(1,1)}_\a, \th^{(1,-1)}_\a, \th^{(0,0)}_\a),\\
&&x^m_\A=x^m+i\th^{(1,1)}\g^m\th^{(-1,-1)}+i\th^{(1,-1)}\g^m\th^{(-1,1)},\nn
\\
&&\d_\eps x^m_\A=-i\eps^{(0,0)}\g^m\th^{(0,0)}-2i\eps^{(-1,1)}\g^m\th^{(1,-1)} 
-2i\eps^{(-1,-1)}\g^m\th^{(1,1)}, 
\eea 
where $\eps^{K\a}=\eps^\a_a U^K_a$ are the harmonic projections of 
the supersymmetry parameters.

General superfields in the analytic coordinates depend also on 
additional spinor coordinates $\th^{(-1,1)}_\a$ and $\th^{(-1,-1)}_\a$.
 The harmonized partial spinor derivatives are
\bea 
&&\pa^{(-1,-1)}_\a=\pa/\pa\th^{(1,1)\a},\q \pa_\a^{(-1,1)}=
\pa/\pa\th^{(1,-1)\a},\q\pa^{(0,0)}_\a=\pa/\pa\th^{(0,0)\a},\lb{partspin}\\
&&\pa^{(1,1)}_\a=\pa/\pa\th^{(-1,-1)\a},\q\pa^{(1,-1)}_\a=\pa/\pa
\th^{(-1,1)\a}.\nn 
\eea

Ordinary complex conjugation connects harmonics of the opposite charges 
\be 
\overline{U_a^{(1,1)}}=U_a^{(-1,-1)},\q\overline{U_a^{(1,-1)}}=U_a^{(-1,1)},
\q \overline{U_a^{(0,0)}}=U_a^{(0,0)}. 
\ee 
We use  the combined conjugation $\sim~$ in the harmonic superspace 
\bea 
&&\widetilde{U^{(p,q)}_a}=U^{(p,-q)}_a,\q 
\widetilde{\th^{(p,q)}_\a}=\th^{(p,-q)}_\a,
\q\widetilde{x^m_\A}=x^m_\A,\nn\\
&&(\th^{(p,q)}_\a\th^{(s,r)}_\b)^\sim=\th^{(s,-r)}_\b\th^{(p,-q)}_\a,\q 
\widetilde{f(x_\A)}=\bar{f}(x_\A), 
\eea 
where $\bar{f}$ is the ordinary complex conjugation. The analytic superspace is 
real with respect to the combined conjugation.

One can  define the combined conjugation for the harmonic 
derivatives of superfields 
\bea 
&&(\pa^{(\pm1,1)}A)^\sim= \pa^{(\pm1,-1)}\tilde{A},\q 
(\pa^{(\pm1,-1)}A)^\sim=\pa^{(\pm1,1)}\tilde{A},\nn\\
&&(\pa^{(\pm2,0)}A)^\sim=-\pa^{(\pm2,0)}\tilde{A},\q 
(\pa^{(0,\pm2)}A)^\sim=\pa^{(0,\mp2)}\tilde{A}. 
\eea

The analytic-superspace integral measure contains partial 
spinor derivatives \p{partspin} 
\bea 
&&d\mu^{(-4,0)}=-\frac{1}{64}dU d^3x_\A 
(\pa^{(-1,-1)})^2(\pa^{(-1,1)})^2(\pa^{(0,0)})^2=
dUd^3x_\A d^6\th^{(-4,0)},\lb{muanal}\\
&&\int  d^6\th^{(-4,0)}(\th^{(1,1)})^2(\th^{(1,-1)})^2(\th^{(0,0)})^2=1. 
\nn 
\eea 
It is pure imaginary 
\be 
(d\mu^{(-4,0)})^\sim=-d\mu^{(-4,0)},\q(d^6\th^{(-4,0)})^\sim=-d^6\th^{(-4,0)}. 
\ee

The harmonic derivatives of the analytic basis commute with the 
generators of the $N{=}5$ supersymmetry 
\bea 
&&\cD^{(1,1)} =\pa^{(1,1)}-i\th^{(1,1)}_\a\th^{(0,0)}_\b\pa^\ab-
\th^{(0,0)\a}\pa^{(1,1)}_\a+\th^{(1,1)\a}\pa^{(0,0)}_\a,\nn\\
&&\cD^{(1,-1)}=\pa^{(1,-1)}-i\th^{(1,-1)}_\a\th^{(0,0)}_\b\pa^\ab-
\th^{(0,0)\a}\pa^{(1,-1)}_\a+\th^{(1,-1)\a}\pa^{(0,0)}_\a=
-(\cD^{(1,1)})^\dagger,\nn\\
&&\cD^{(2,0)}=[\cD^{(1,-1)},\cD^{(1,1)}] 
=\pa^{(2,0)}-2i\th^{(1,1)}_\a\th^{(1,-1)}_\b\pa^\ab-\th^{(1,-1)\a}\pa^{(1,1)}_\a
+\th^{(1,1)\a}\pa^{(1,-1)}_\a,\nn\\
&&\cD^{(0,2)}=\pa^{(0,2)}+\th^{(1,1)\a}\pa^{(-1,1)}_\a
-\th^{(-1,1)\a}\pa^{(1,1)}_\a\nn\\
&& \cD^{(0,-2)}=-(\cD^{(0,2)})^\dagger=\pa^{(-2,0)} 
+\th^{(1,-1)\a}\pa^{(-1,-1)}_\a-\th^{(-1,-1)\a}\pa^{(1,-1)}_\a.\nn 
\eea 
Note that   harmonic derivatives $\cD^{(0,\pm2)}$ change 
the second U(1) charge; these operators do not act on $x^m_\A$, 
in distinction with other harmonic derivatives. It is useful to 
define the AB-representation of the U(1) charge operators 
\bea 
&&\cD^0_1 A^{(p,q)}=p\,A^{(p,q)},\q \cD^0_2 
A^{(p,q)}=q\,A^{(p,q)}, 
\eea 
where $A^{(p,q)}$ is an arbitrary harmonic superfield in AB.

The spinor derivatives  in the analytic basis are 
\bea 
&&D^{(-1,-1)}_\a=\pa^{(-1,-1)}_\a+2i\th^{(-1,-1)\b}\pa_\ab,\q 
D^{(-1,1)}_\a=\pa^{(-1,1)}_\a+2i\th^{(-1,1)\b}\pa_\ab,\nn\\
&& D^{(0,0)}_\a=\pa^{(0,0)}_\a+i\th^{(0,0)\b}\pa_\ab,\q 
D^{(1,1)}_\a=\pa^{(1,1)}_\a,\q D^{(1,-1)}_\a=\pa^{(1,-1)}_\a. 
\eea

The analytic superfields $\L(\zeta,U)$ depend on harmonics and 
the analytic coordinates, and satisfy the Grassmann analyticity 
conditions 
\be G:\qq D^{(1,\pm1)}_\a\L=0.\lb{GA} 
\ee 
The action of the five harmonic derivatives \be \cD^{(1,\pm1)},\q 
\cD^{(2,0)},\q\cD^{(0,\pm2)}\lb{5hder} 
\ee 
preserves this $G$-analyticity.

 \setcounter{equation}0
\section{Chern-Simons model in $N=5$ analytic superspace}

Using the harmonic-superspace method \cite{GIOS} we introduce 
the  $D{=}3, N{=}5$ analytic matrix gauge prepotentials 
corresponding to the five harmonic derivatives \p{5hder} 
\bea
&&V^{(p,q)}(\z,U)=[V^{(1,1)},\q V^{(1,-1)},\q V^{(2,0)},\q V^{(0,\pm2)}],
\nn\\
&&(V^{(1,1)})^\dagger=-V^{(1,-1)},\q 
(V^{(2,0)})^\dagger=V^{(2,0)},\q 
V^{(0,-2)}=[V^{(0,2)}]^\dagger, \lb{real} 
\eea 
where the Hermitian conjugation $\dagger$ includes $\sim$ conjugation of 
matrix elements and transposition. The infinitesimal gauge 
transformations of these prepotentials depends on the analytic 
anti-Hermitian matrix gauge parameter $\L$ 
\bea 
&&\d_\L V^{(1,\pm1)}=\cD^{(1,\pm1)}\L+[V^{(1,\pm1)},\L],\q
\d_\L V^{(2,0)}=\cD^{(2,0)}\L+[V^{(2,0)},\L],\nn\\
&&\d_\L V^{(0,\pm2)}=\cD^{(0,\pm2)}\L +[V^{(0,\pm2)},\L]. 
\eea

We shall consider the restricted gauge supergroup using the 
supersymmetry-preserving harmonic ($H$) analyticity  
constraints on the gauge superfield parameters 
\be 
H1:\qq \cD^{(0,\pm2)}\L=0.\lb{LHA} 
\ee 
These constrains yield  additional reality conditions for the component gauge 
parameters.

We use in this paper  the harmonic-analyticity constraints on 
the gauge prepotentials 
\be 
H2:\qq V^{(0,\pm2)}=0,\q 
\cD^{(0,-2)}V^{(1,1)}=V^{(1,-1)}, \q \cD^{(0,2)}V^{(1,1)}=0 
\ee 
and the conjugated constraints  combined with relations 
\p{real}. It is evident that the $G$- and $H$-analyticities of 
the prepotentials are preserved by the restricted gauge 
transformations \p{LHA}. 

Now we have only three gauge prepotentials in complete analogy 
with the algebraic structure of the gauge theory in the $N{=}3, 
D{=}4$ harmonic superspace \cite{GIOS}. The superfield CS action 
can be constructed in terms of these $H$-constrained gauge superfields 
\bea 
&&S=-\frac{2i}{3g^2}\int   d\mu^{(-4,0)}\Tr\{V^{2,0}\cD^{(1,-1)}
V^{(1,1)}+V^{1,1}\cD^{(2,0)}V^{(1,-1)}\nn\\
&&+V^{1,-1}\cD^{(1,1)}V^{(2,0)}+V^{2,0}[V^{(1,-1)},V^{(1,1)}]-\sfrac12 
V^{(2,0)}V^{(2,0)}\},\lb{CSact} 
\eea 
where $g$ is the dimensionless CS  coupling constant.

The corresponding superfield  gauge equations of motion have 
the following form: 
\bea 
&&F^{3,-1}=\cD^{(1,-1)}V^{(2,0)}-\cD^{(2,0)}V^{(1,-1)}+[V^{(1,-1)},V^{(2,0)}]=0,
\nn\\
&&F^{3,1}=\cD^{(1,1)}V^{(2,0)}-\cD^{(2,0)}V^{(1,1)}+[V^{(1,1)},V^{(2,0)}]=0,
\lb{CSequ}\\
&&V^{(2,0)}=\cD^{(1,-1)}V^{(1,1)}-\cD^{(1,1)}V^{(1,-1)}+[V^{(1.-1)},V^{(1,1)}]
\equiv \hat{V}^{(2,0)}.\lb{Vcomp} 
\eea

The last prepotential can be composed algebraically in terms of 
two other basic superfields. Using the substitution  
$V^{(2,0)}\rightarrow \hat{V}^{(2,0)}$ in \p{CSact} we can 
obtain the  alternative form of the  action with only two 
independent prepotentials $V^{1,1}$ and $V^{1,-1}$ 
\bea 
&&S_2=-\frac{2i}{3g^2}\int  
d\mu^{(-4,0)}\Tr\{V^{1,1}\cD^{(2,0)}V^{(1,-1)}\nn\\
&&+\sfrac12\left(\cD^{(1,-1)}V^{(1,1)}-\cD^{(1,1)}V^{(1,-1)}+[V^{(1,-1)},
V^{(1,1)}] \right)^2\}.\lb{CSact2} 
\eea

It is evident that the superfield action \p{CSact} is invariant 
with respect to the sixth supersymmetry transformation defined 
on our gauge prepotentials 
\bea 
\d_6[V^{(1,\pm1)},V^{(2,0)}]=\eps^\a_6D^{(0,0)}_\a[V^{(1,\pm1)},V^{(2,0)}], 
\eea 
where $\eps^\a_6$ are the corresponding spinor  
parameters. Thus, our superfield gauge model possesses the 
$D{=}3, N{=}6$ supersymmetry.

The $D{=}3, N{=}5$ superconformal transformations can be defined on 
the analytic coordinates. For instance,  the special conformal 
$K$-transformations are 
\bea 
&&\d_k x^\ab_\A=
\sfrac12x^{\a\g}_\A\,k_{\g\r} x^{\b\r}_\A+2l\,x^\ab_\A,\nn\\
&&\d_k\th^{(0,0)\a}=\sfrac12x^{\a\b}_\A\th^{(0,0)\g}k_{\b\g},\\
&&\d_k\th^{(1,1)\a}= \sfrac12x^{\a\b}_\A\th^{(1,1)\g}k_{\b\g} 
+\sfrac{i}4\tooS\th^{(1,1)\b}k^\a_\b, 
\eea 
where $k_\ab=k^m(\g_m)_\ab$ are the corresponding parameters. The 
$K$-transformations of the harmonics have the form 
\bea 
&&\d_k U^{(0,0)}_b=-\l^{(1,1)}_kU^{(-1,-1)}_b-\l^{(1,-1)}_kU^{(-1,1)}_b,\q
\d_k U^{(1,1)}_b=\l^{(1,1)}_kU^{(0,0)}_b+\l^{(2,0)}_kU^{(-1,1)}_b,\nn\\
&&\d_k U^{(1,-1)}_b=\l^{(1,-1)}_kU^{(0,0)}_b-\l^{(2,0)}_kU^{(-1,-1)}_b,\q
\d_k U^{(-1,\pm1)}_b=0,\\
&& \l^{(1,1)}_k=ik_\ab\th^{(1,1)\a}\th^{(0,0)\b},\q 
\l^{(1,-1)}_k=ik_\ab\th^{(1,-1)\a}\th^{(0,0)\b},\q 
\l^{(2,0)}_k=ik_\ab\th^{(1,1)\a}\th^{(1,-1)\b}.\nn 
\eea 
The special supersymmetry transformations of all coordinates can be 
obtained via the Lie bracket $\d_\eta=[\d_\eps,\d_k]$.

It is easy to check that the analytic integral measure 
$\mu^{(-4,0)}$ \p{muanal} is invariant with respect to these 
superconformal transformations.

The special conformal transformations of the harmonic 
derivatives have the following form: 
\bea 
&&\d_k\cD^{(1,1)}=-\sfrac12\l^{(1,1)}_k(\cD^0_1+\cD^0_2)-\l^{(1,-1)}_k\cD^{(0,2)},
\nn\\
&&\d_k\Dpm=-\sfrac12\lpm_k(\cD^0_1-\cD^0_2)-\lpp_k\DoM,\nn\\
&&\d_k\DPo=\lpo_k\cD^0_2+\lpm_k\Dpp-\lpp_k\Dpm,\lb{SCH}\\
&&\d_k\DoP=\d_k\DoM=0,\q \d_k\cD^0_1=\d_k\cD^0_2=0,\nn 
\eea 
and the SO(5) and special supersymmetry transformations can be 
defined analogously.

The $K$-transformations of the gauge prepotentials are 
\bea
&&\d_k V^{(1,1)}=0,\q\d_k V^{(1,-1)}=0,\nn\\
&&\d_k V^{(2,0)}=\lpm_k V^{(1,1)}-\lpp_k V^{(1,-1)}=\d_k 
\hat{V}^{(2,0)}, 
\eea 
where $\hat{V}^{(2,0)}$ is the composite prepotential \p{Vcomp}.

It is easy to check directly the superconformal invariance of 
the gauge actions $S$ \p{CSact} and $S_2$  \p{CSact2}.

The classical superfield equations \p{CSequ} and \p{Vcomp} have 
only pure gauge solution 
\bea 
V^{(1,\pm1)}=e^{-\L}\cD^{(1,\pm1)}e^\L,\q 
V^{(2,0)}=e^{-\L}\cD^{(2,0)}e^\L, \lb{pure} 
\eea 
where $\L$ is an arbitrary anti-Hermitian matrix superfield satisfying the 
conditions \p{LHA}.

\setcounter{equation}0
\section{Harmonic component fields in the $N=6$  Chern-Simons model}

Let us consider the U(1) gauge group. The pure gauge degrees of 
freedom in the abelian prepotential $V^{(1,1)}$ can be 
eliminated by the transformation $\d V^{(1,1)}=\cD^{(1,1)}\L$. 
In the WZ-gauge we have $\L_{WZ}=ia(x_\A)$. The harmonic 
decomposition of the $HA$-constrained prepotential $V^{(1,1)}$ 
in the WZ-gauge has the following form: 
\bea 
&&V^{(1,1)}_{WZ}=V^{(1,1)}_0+V^{(1,1)}_1+O(U^2),\q 
\cD^{(0,-2)}V^{(1,1)}_{WZ}=-[V^{(1,1)}_{WZ}]^\sim,\nn\\
&&
V^{(1,1)}_0=(\th^{(1,1)}\g^m\th^{(0,0)})A_m+i(\th^{(0,0)})^2\th^{(1,1)\a}\j_\a,\\
&&V^{(1,1)}_1=(\th^{(0,0)})^2U^{(1,1)}_aB_a 
+(\th^{(1,1)}\g^m\th^{(1,-1)})U^{(-1,1)}_aC^a_m\nn\\
&&+i\th^{(1,1)(\a}\th^{(1,-1)\b}\th^{(0,0)\g)}U^{(-1,1)}_a\J^a_{\a\b\g}
+i(\th^{(1,-1)}\th^{(0,0)})\th^{(1,1)\a}U^{(-1,1)}_a\xi^a_\a\nn\\
&&-i(\th^{(1,1)}\th^{(0,0)})\th^{(1,-1)\a}U^{(-1,1)}_a\xi^a_\a
+i(\th^{(1,1)})^2(\th^{(0,0)})^2U^{(-1,-1)}_aR^a\nn\\
&& +i(\th^{(1,1)}\th^{(1,-1)})(\th^{(0,0)})^2U^{(-1,1)}_a R^a 
+i(\th^{(1,1)}\g^m\th^{(1,-1)})(\th^{(0,0)})^2U^{(-1,1)}_aG^a_m, 
\eea 
where all terms are parametrized by the real off-shell 
bosonic  fields $A_m, B_a, C^a_m, R^a$ and $ G^a_m$ or the real 
Grassmann fields $\j_\a, \J^a_{\a\b\g}$ and $\xi^a_\a$. The 
higher harmonic terms in $V^{(1,1)}_{WZ}$ contain an infinite 
number of the SO(5) tensor fields. In the  gauge group SU(n), 
all component fields are the Hermitian traceless matrices.

Two other U(1) prepotentials contain the same component fields
in the WZ-gauge 
\bea
&&V^{(1,-1)}_{WZ}=\cD^{(0,-2)}V^{(1,1)}_{WZ}=V^{(1,-1)}_0+V^{(1,-1)}_1+
O(U^2),\\
&&\hat{V}^{(2,0)}_{WZ}=\cD^{(1,-1)}\Vpp_{WZ}-\cD^{(1,-1)}\Vpm_{WZ}=
V^{(2,0)}_0+ V^{(2,0)}_1+O(U^2).\nn 
\eea

The superfield terms
$$\Tr[\Vpp_0\DPo\Vpm_0+\sfrac12(\Dpm\Vpp_0-\Dpp\Vpm_0+[\Vpm_0,\Vpp_0])^2]
$$
in the action $S_2$ \p{CSact2} yield the following contribution to 
the  component Lagrangian: 
\be 
L_0=\ve^{mnr}\Tr A_m(\pa_nA_r+\sfrac{i}3[A_n,A_r])-\sfrac{i}3\Tr\j^\a\j_\a. 
\ee 
The superfield terms 
\bea 
&&\Tr\{\Vpp_1\DPo\Vpm_1+\sfrac12(\Dpm\Vpp_1-\Dpp\Vpm_1+[\Vpm_0,\Vpp_1]
+[\Vpm_1,\Vpp_0])^2\}\nn\\
&&+\Tr \{(\Dpm\Vpp_0-\Dpp\Vpm_0)[\Vpm_1,\Vpp_1]\} 
\eea 
give us the  Lagrangian for the SO(5) vector fields 
\bea 
&&L_1=\sfrac{2}{5}\Tr C^a_m(\pa^mB_a+i[A_m,B_a])-\sfrac{8}{15}\Tr C^a_mG^m_a 
-\sfrac45\Tr B_aR_a \nn\\&& +\sfrac{i}6\Tr \xi^\a_a\xi_{a\a} 
-\sfrac{i}{120}\Tr\J^{a\a\b\g}\J^a_{\a\b\g}. 
\eea 
It is not difficult to construct the component Lagrangian for the SO(5) 
tensor fields.

The $N{=}6$ CS equations of motion for the lowest SO(5) 
component fields are 
\be 
\ve^{mnr}(\pa_nA_r-\pa_rA_n+i[A_n,A_r])=0,\q 
\j_\a=C^a_m=B_a=R_a=G^a_m=\xi^a_\a= \J^a_{\a\b\g}=0. 
\ee 
All SO(5) tensor auxiliary fields also vanish on-shell. The 
superfield representation of this pure gauge solution in the 
WZ-gauge is 
\be 
V^{(1,\pm1)}_{WZ}=e^{-ia}\cD^{(1,\pm1)}e^{ia},\q 
V^{(2,0)}_{WZ}=e^{-ia}\cD^{(2,0)}e^{ia}. 
\ee

\section{Conclusion and acknowledgements}
We considered the superfield  model with the $D{=}3, 
N{=}6$ superconformal supersymmetry. The action of this model is 
constructed in the $N{=}5$ harmonic superspace using the  
Grassmann and harmonic analyticity conditions. The classical superfield
equations of motions for the analytic Chern-Simons gauge prepotentials 
have the pure gauge solution only. In the field-component representation, 
the action of this model contains the Chern-Simons term for the vector 
gauge field and an infinite number of the interaction terms for the 
auxiliary bosonic and fermionic fields. All auxiliary fields vanish 
on-shell. The superfield representation is useful for the quantization 
and perturbative calculations.

{\bf Note added in proof.} P.S. Howe informed me that the harmonic-superspace
description of the $N{=}5, 6$ Chern-Simons theories was considered in the paper
\cite{HL}. It should be stressed that our harmonic constraints
\p{LHA} reduce the SO(5)/U(1)$\times$U(1) space to the SO(5)/U(2) space
proposed in \cite{HL}. 

I am grateful to E.A. Ivanov for  interesting discussions and to P.S. Howe for
the important comments. This 
work was partially supported by  DFG grant 436 RUS 113/669-3 , 
by RFBR grants 06-02-16684 and 06-02-04012, by  INTAS grant 
05-10000008-7928 and by  grant of the Heisenberg-Landau 
programme.

\end{document}